\documentclass[aps,pre,twocolumn,showpacs]{revtex4-1}

\usepackage{graphicx}

\begin{document}

\title{Pinning time statistics for vortex lines in disordered environments}

\author{Ulrich Dobramysl}
\affiliation{Mathematical Institute, University of Oxford, 
  Oxford OX2 6GG, U.K.}
\author{Michel Pleimling}
\affiliation{Department of Physics (MC 0435), Virginia Tech, 
  Blacksburg, VA 24061, USA}
\author{Uwe C. T\"auber}
\affiliation{Department of Physics (MC 0435), Virginia Tech, 
  Blacksburg, VA 24061, USA}

\begin{abstract}
  We study the pinning dynamics of magnetic flux (vortex) lines in a disordered
  type-II superconductor. Using numerical simulations of a directed elastic line
  model, we extract the pinning time distributions of vortex line segments. We
  compare different model implementations for the disorder in the surrounding
  medium: discrete, localized pinning potential wells that are either attractive
  and repulsive or purely attractive, and whose strengths are drawn from a
  Gaussian distribution; as well as continuous Gaussian random potential
  landscapes. We find that both schemes yield power law distributions in the
  pinned phase as predicted by extreme-event statistics, yet they differ
  significantly in their effective scaling exponents and their short-time
  behavior.
\end{abstract}




\maketitle

\section{Introduction}
\label{sec:introduction}

The static and dynamic properties of elastic manifolds in random media have 
been central research topics of statistical physics for decades (see, e.g., 
Refs.~\cite{Fisher1998, Brazovskii2004}). Specifically, fluctuating directed 
lines interacting with spatially uncorrelated disorder represent the basic 
model for magnetic vortices in type-II superconductors with point pinning 
centers~\cite{Blatter1994}, crystal dislocations~\cite{Ioffe1987}, as well as 
for aligned polymers~\cite{Ertas1992, Ertas1996, Kardar1998}. There are also 
fascinating intimate mathematical connections with the dynamics of driven
interfaces and non-equilibrium growth processes~\cite{HalpinHealy1995, 
Rosso2003}. A system of directed lines, subject to competing thermal 
fluctuations and pinning from a random disorder background, constitutes a 
remarkably complex system displaying a rich thermodynamic phase diagram and a 
wealth of distinct dynamical regimes~\cite{Blatter1994,Agoritsas2012}, as well 
as intriguing non-equilibrium relaxation kinetics~\cite{Du2007, Pleimling2011, 
Dobramysl2013, Assi2014}. In particular, driven elastic strings in a random 
medium show a transition between a pinned vortex glass phase, in which the 
dynamics are dominated by thermally activated creep, and a flowing phase above 
a critical depinning force~\cite{Nattermann2000}. Both phases possess rich 
dynamical features~\cite{Duemmer2005, Kolton2009}, with universal depinning 
force distributions~\cite{Bolech2004} and scaling 
behavior~\cite{Bustingorry2010} at the critical point.

In this work, we focus on the statistical distribution of dwelling times of 
line segments localized at the defects, which incorporates information on the 
collective Larkin--Ovchinnikov pinning scale. We perform detailed computer 
simulations to investigate how distinct model representations of the 
disordered environment affect the depinning kinetics, and compare our 
numerical data with the hitherto unconfirmed theoretical predictions of 
Ref.~\cite{Vinokur1996}. We begin by introducing our model Hamiltonian and 
providing key theoretical background. We then describe our dynamical simulation 
method based on an overdamped Langevin equation. We next discuss different 
disorder implementations, our simulation protocol, and the measurement 
procedures we employ in order to extract pinning time distributions. We then 
proceed with a brief analysis of the effects of overlapping disorder 
potentials, and the transition between the pinned and free-flowing phases. Our 
principal results concern the dwelling time statistics from simulations with 
either discrete, localized pinning potential wells, or with smooth Gaussian 
pinning landscapes. 

\section{Theoretical Background}
\label{sec:theor-backgr}

We consider non-interacting (independent) vortex lines driven through a 
three-dimensional disordered superconductor, with the orienting magnetic field
aligned with the $z$ direction. Line segments can move in the perpendicular 
$xy$ plane, but cannot form loops. The corresponding coarse-grained 
Hamiltonian for the directed lines is given by~\cite{Nelson1993, Blatter1994}
\begin{equation}
  \label{eq:hamiltonian}
  \mathcal{H}[\vec{r}] = \int_0^{\gamma  L} \biggl[ \frac{\epsilon}{2}
  \biggl| \frac{d\vec{r}(z)}{dz} \biggr|^2 + U\bigl( \vec{r}(z) , z \bigr) 
  \biggr] dz \ .
\end{equation}
It constitutes the effective energy functional of the two-dimensional vector
$\vec{r}(z)$ that defines the line position along the line axis $z$, i.e. the
line trajectory at a given time. The requirement that the directed elastic lines
may not form loops is reflected in the condition that $\vec{r}(z)$ be
surjective. The line tension $\epsilon$ is the elastic energy per unit length,
and $U\bigl( \vec{r}(z) , z \bigr)$ represents the disorder potential. Its
detailed implementation for either discrete pinning sites or a smooth potential
landscape will be described below. To suppress surface effects, we employ
periodic conditions along the $z$ direction (identifying $z = \gamma L$ with 
$z = 0$).

In order to capture the dynamics of our vortex system, we employ overdamped 
Langevin dynamics~\cite{Brass1989, Dobramysl2013}:
\begin{equation}
  \label{eq:langevin_eq}
  \eta \, \frac{\partial\vec{r}(z,t)}{\partial t} =
  - \frac{\delta \mathcal{H}[\vec{r}]}{\delta \vec{r}(z,t)} 
  + \vec{f}(z,t) + \vec{F} \ .
\end{equation}
Here, $\eta$ is the viscosity of the surrounding medium. For magnetic flux
lines, it is given by the Bardeen--Stephen viscous drag 
parameter~\cite{Bardeen1965}. The constant force density vector $\vec{F}$ 
represents the external drive. Fast, microscopic degrees of freedom stemming 
from interactions with the surrounding medium are captured via thermal 
stochastic forcing, modeled as uncorrelated Gaussian white noise with zero 
mean $\langle \vec{f}(z,t) \rangle = 0$ and the second moment 
$\langle f_\alpha(z,t) f_\beta(z',t') \rangle = 2 \eta T \delta_{\alpha \beta}
 \, \delta(z-z') \delta(t-t')$ ($\alpha,\beta = x,y$), satisfying Einstein's 
relation for thermal equilibrium at temperature $T$ (we set Boltzmann's 
constant $k_{\rm B} = 1$). 

The vortices are thus subject to various energy scales: (i) the internal 
elastic energy, (ii) the (random) disorder potential, (iii) an external 
driving force, and (iv) thermal fluctuations stemming from interactions with 
the surrounding medium. Varying the strengths of these competing contributions 
leads to remarkably rich and complex dynamics. At $T = 0$, there exists a 
sharp continuous transition at a critical driving force $F_c$ separating a 
pinned vortex phase from a non-equilibrium steady state in which the lines are 
freely flowing~\cite{Blatter1994, Vinokur1996, Fisher1998}. At finite 
temperatures, the dynamic phase transition is thermally rounded, resulting in
line motion (flux creep) even below the critical depinning 
force~\cite{Fisher1998}. Depending on the spatial distribution and strength of 
the disorder, the transverse line roughness displays intriguing behavior near 
$F_c$~\cite{Dobramysl2013}.

The relevant length scale in the pinned phase is given by the
Larkin--Ovchinnikov pinning length $L_L = \xi (\epsilon / w)^{2/3}$, with $\xi$ 
and $w$ denoting the spatial range of disorder correlations and the standard 
deviation of the disorder potential strength. $L_L$ measures the typical 
extent of collectively pinned line segments~\cite{Vinokur1996}. The transition 
to the free-flowing phase occurs when the driving force becomes large enough 
to cause displacements on the order of $L_L$. Associated with this length 
scale is a minimum energy barrier between pinned configurations 
$E_L \approx \epsilon \xi^2 / L_L$. Thus one may estimate the critical force 
(per line element length) as 
$F_c \approx E_L / (\xi L_L) \approx w^{4/3} / (\xi \epsilon^{1/3})$. Using 
arguments based on extreme-event statistics, Vinokur, Marchetti and Chen found 
that the pinning time distribution of line segments should obey a power law 
for large dwell times $\tau$,
\begin{equation}
  \label{eq:powerlaw}
  P(\tau) \propto \tau^{-1-\alpha} \ , \quad \alpha \propto T / E_L \ ,
\end{equation}
with a scaling exponent $\alpha < 1$ for low temperatures~\cite{Vinokur1996}. 
In their derivation, Vinokur {\em et al.} assumed a Gaussian-distributed 
disorder potential with a spatial correlation length $\xi$. However, material 
defects in superconducting samples should more realistically be represented by 
discrete and moreover purely attractive pinning sites instead of a continuous 
disordered landscape with zero mean. We remark that studies of non-equilibrium 
vortex relaxation kinetics have emphasized the drastic influence of the 
underlying pinning model~\cite{Iguain2009, Pleimling2011}. At any rate, the
intriguing theoretical prediction (\ref{eq:powerlaw}) has not yet been
numerically tested (nor experimentally confirmed).

\section{Model Description}
\label{sec:model}

\subsection{Discretized elastic line model}

In order to facilitate computational modeling of this system, we discretize 
the elastic line into connected nodes with a spacing $\gamma$ along the $z$ 
axis. Line tension is implemented via an elastic interaction between adjacent 
nodes. The ensuing discretized Hamiltonian reads
\begin{equation}
  \label{eq:disc_hamiltonian}
  H(\{ \vec{r}_k \}) = \sum_{k=1}^L \biggl[ \frac{\epsilon}{4 \gamma}
  |\vec{r}_{k-1} + \vec{r}_{k+1} - 2 \vec{r}_k|^2 
  + \gamma U(\vec{r}_k , k \gamma) \biggr] \ ,
\end{equation}
where $k = 0$ maps to $k = L$ in order to correctly account for the periodic 
boundaries in the $z$ direction. We now have to numerically solve the coupled
Langevin equations 
\begin{equation}
  \label{eq:disc_langevin_eq}
  \gamma \eta \, \frac{\partial \vec{r}_j(t)}{\partial t} =
  - \frac{\partial H(\{ \vec{r}_k \})}{\partial \vec{r}_j(t)} + \vec{f}_j(t) 
  + \gamma \vec{F}
\end{equation}
with $\vec{f}_j(t) = \gamma \vec{f}(z,t)$; following Ref.~\cite{Brass1989}, we 
perform the temporal integration via the Euler--Maruyama method.

In the following, all lengths are measured in terms of the pinning center
radius $b_0=3.5\,\text{nm}$, and energies relative to the intrinsic vortex 
line energy $\epsilon_0 b_0$. Inserting parameter values that correspond to 
the high-$T_c$ superconducting compound YBCO, one obtains $\epsilon_0 \approx 
1.92 \times10^{-6} \, \text{erg} / \text{cm}$~\cite{Pleimling2011}. The vortex 
line tension in this anisotropic material is 
$\epsilon \approx 0.189 \epsilon_0$, whence the Bardeen--Stephen viscous drag 
coefficient becomes 
$\eta \approx 10^{-10} \text{erg} \times \text{s} / \text{cm}^3$, which yields 
the fundamental simulation time unit 
$t_0 \approx 18\,\text{ps}$~\cite{Dobramysl2013}. We set the layer spacing 
equal to the pinning center radius $\gamma = b_0$.
In previous work~\cite{Dobramysl2013, Assi2014}, we have extensively tested 
this discretized elastic line model, and verified that it correctly reproduces
the expected thermodynamic phases (for $\vec{F} = 0$) as well as the 
established non-equilibrium steady-state properties at finite drive current.

\subsection{Disorder potential}
\label{sec:disorder_pot}

We investigate and compare fundamentally different disorder implementations: 
discrete potential wells with varying strengths, and Gaussian distributed 
potential landscapes with a finite correlation length. The former scheme
constitutes a more realistic model of localized pinning sites for flux lines 
in type-II superconductors \cite{Pleimling2011, Dobramysl2013}, while the 
latter is specifically amenable to analytic investigations such as 
Ref.~\cite{Vinokur1996}.

\subsubsection{Discrete pinning sites}
\label{sec:discrete_pins_model}

In type-II superconductors, material defects such as oxygen vacancies take 
the form of randomly distributed discrete potential wells. They act as 
short-ranged pinning centers wherein magnetic flux lines may be trapped. (In 
this work, we only consider uncorrelated point-like disorder.) We model these 
individual pinning sites as an ensemble of $N_k$ smooth, radially symmetric 
potential wells per layer $k$, centered at ${\vec r}^{\,(i)}_k$:
\begin{equation}
  \label{eq:discrete_pinning}
  U(\vec{r}_k , k \gamma) = \sum_{i=1}^{N_k} \frac{p^{(i)}_k}{2} \Bigl( 1 - 
  \tanh \Bigl[ 5 \bigl( \big| \vec{r}_k - \vec{r}^{\,(i)}_k \big| / b_0 - 1 
  \bigr) \Bigr] \Bigr) \ .
\end{equation}
The pinning potential strengths $p^{(i)}_k$ are Gaussian random variables 
with mean $\big\langle p^{(i)}_k \big\rangle = \mu$ and variance 
$\big\langle p^{(i)}_k p^{(j)}_l \big\rangle = w^2 \delta_{ij} \delta_{kl}$. 
The pin positions $\vec{r}^{\,(i)}_k$ ($i = 1, \ldots, N_k$) are uniformly 
distributed throughout the $xy$ domain, independently in each layer $k$. 
Overlaps between pinning sites are avoided, hence the minimal distance between
sites is $\big| \vec{r}^{\,(i)}_k - \vec{r}^{\,(j)}_l \big| > 2 b_0$. (The 
effects of overlapping defect potentials in the depinned phase will be 
discussed below.)

\subsubsection{Continuous disorder landscape}
\label{sec:landscape_pinning}

Alternatively, we employ a Gaussian pinning potential landscape in order to 
connect to the model considered by Vinokur {\it et al.} \cite{Vinokur1996}. 
To generate a continuous smooth disorder landscape, we draw a potential value 
from a Gaussian distribution at each node of a square lattice with spacing 
$\xi$. Such a lattice is constructed independently for each layer $k$. The 
value of the disorder potential at an arbitrary point $(\vec{r}_k, k \gamma)$ 
is then determined via a bilinear interpolation of the values defined on the 
lattice nodes. The potential landscape $U(\vec{r},z)$ resulting from this 
procedure is characterized by correlations on the order of the lattice spacing
$\xi$: $\big\langle U(\vec{r}, k \gamma) U(\vec{r\,}', l \gamma) \big\rangle =
 w^2 \delta_{kl} f\bigl( |\vec{r} - \vec{r\,}'| / \xi \bigr)$ with a (roughly
exponentially) decaying function $f$, similar to Ref.~\cite{Vinokur1996}. In
comparisons with discrete localized pinning potentials, we set $\xi = b_0$.

\section{Simulation procedure}
\label{sec:simulation-procedure}

\subsection{Initialization}

We initialize a simulation by creating a computational domain of length $l = 100
b_0$ (along the direction of the driving force), width $W = 10 b_0$, and height
$\gamma L$, with $L = 200$ and periodic boundary conditions in all
directions. We have systematically varied the line length and found no
finite-size effects already at this value of $L$. Depending on the desired type
of disorder, we either randomly distribute discrete pinning centers throughout
the domain, or create a continuous disorder landscape through interpolation from
normal-distributed lattice potential values. The resulting force field is then
used in the numerical solution of Eqs.~(\ref{eq:disc_langevin_eq}).

Ten lines are then introduced at regular intervals along the system length 
$l$. Initially, the line elements are perfectly aligned along the $z$ 
direction. While in the collectively pinned phase, the lines will sample the 
pinning environment only in the vicinity of their starting position during the
simulation time window. Provided their initial mutual distances are 
sufficiently large, they will never overlap. We may therefore introduce 
multiple vortex lines in order to improve our computational efficiency. The 
initially straight vortices are allowed to relax towards a driven 
non-equilibrium steady state. During this simulation phase, thermal 
fluctuations cause transverse displacements and the directed lines to roughen. 
After they have assumed a steady-state configuration, data on the pinning time 
distributions are collected by recording discrete depinning events of 
individual line elements. 

\subsection{Measurement procedure}
\label{sec:meas-proc}

In order to measure the distribution of dwelling times for the different
pinning schemes, we need to define respective depinning criteria. In the case
of discrete pinning sites, we consider a line element pinned if its distance 
to the closest defect site is less than the pinning potential's radius 
$b_0$, i.e. if it is located inside the pinning site. To acquire the time a 
line element has spent attached to the pin, we track its position and record 
the instant when it first enters the pinning center. When the line element 
leaves the site, the elapsed time difference $\tau$ is stored as the dwelling 
time.

For a continuous disorder landscape, the choice of a suitable depinning 
criterion is less obvious. After the relaxation phase, we take a snapshot of 
the positions of all line elements. We periodically check the distance each 
line element has moved and probe if its separation from the saved location is 
larger than $2 \xi$. If the line element has left the vicinity of its 
``pinning location'', we record the time $\tau$ it spent there as the pinning 
duration. The line element's new position becomes its updated pinning 
location.

In either situation, we generate histograms from the stored dwelling times. 
We use these to approximate the probability distribution $P(\tau)$ for the 
time $\tau$ that single line elements persist in a pinned configuration.

\subsection{Overlapping discrete disorder}
\label{sec:overlap}

\begin{figure}[t]
  \centering
  \includegraphics[width=0.9\columnwidth]{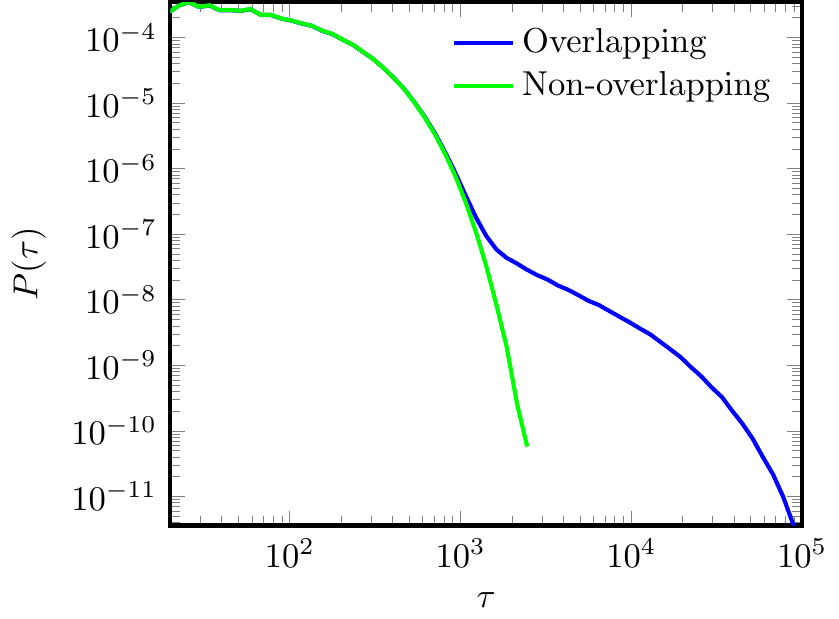} 
  \caption{Influence of overlapping defect potentials on the pinning time 
    distribution for single line elements ($L = 1$) in the free-flowing 
    phase. The blue (dark) graph shows data from simulations in which 
    discrete pinning potential wells were allowed to overlap, while the green
    (light) curve displays results for strictly non-overlapping sites. 
    Simulation parameters: $T = 0.02 \epsilon_0 b_0$, 
    $F = 0.002 \epsilon_0 b_0^{-1}$, $N_k = 50$, $\mu = 0.1 \epsilon_0$, 
    $w = 0.001 \epsilon_0$; the data were averaged over $1000$ independent
    realizations.}
  \label{fig:overlapping}
\end{figure}

Before we investigate the resulting pinning time distributions in detail, we
briefly address the influence of overlapping discrete potential wells. A 
uniformly random positioning of pinning sites inevitably leads to spatially 
overlapping defect potentials. This, of course, generates regions where the 
disorder potential is considerably stronger than for single pins. 
Figure~\ref{fig:overlapping} shows the effects of overlapping sites on 
$P(\tau)$ in the free-flowing phase. The pinning time distribution exhibits
two successive humps if overlapping sites are allowed. The second flat region 
disappears when existing pinning centers are strictly avoided during the 
placement of new sites. We interpret the data for the potentially overlapping 
pins as reflecting a two-step process: When a line element is trapped inside 
an isolated site with typical depth $U_1 \approx \mu$, its escape time is on 
the order of $\tau_1 \sim 10^2 t_0$. When overlaps are prohibited, this is 
the only relevant time scale. Yet in the presence of overlapping wells, there 
exist regions with a disorder potential that, on average, is twice as deep as 
single sites, $U_2 \approx 2 \mu$. Line elements trapped in such deeper 
troughs require a much longer time $\tau_2$ to leave the trap. From 
Fig.~\ref{fig:overlapping}, we infer $\tau_2 \approx \tau_1^2 \sim 10^4 t_0$, 
which is consistent with Kramers' solution for the escape time problem, 
wherein the mean escape time is proportional to $\exp(- U / T)$. In 
Ref.~\cite{Lama2009}, the authors similarly observe multiple plateaus in the 
activity statistics of elastic strings at low temperatures. In the remainder 
of this article, pinning site overlaps are explicitly disallowed.

\section{Simulation Results}

\subsection{Pinned versus free-flowing phase}
\label{sec:phases}

\begin{figure}
  \centering
  \includegraphics{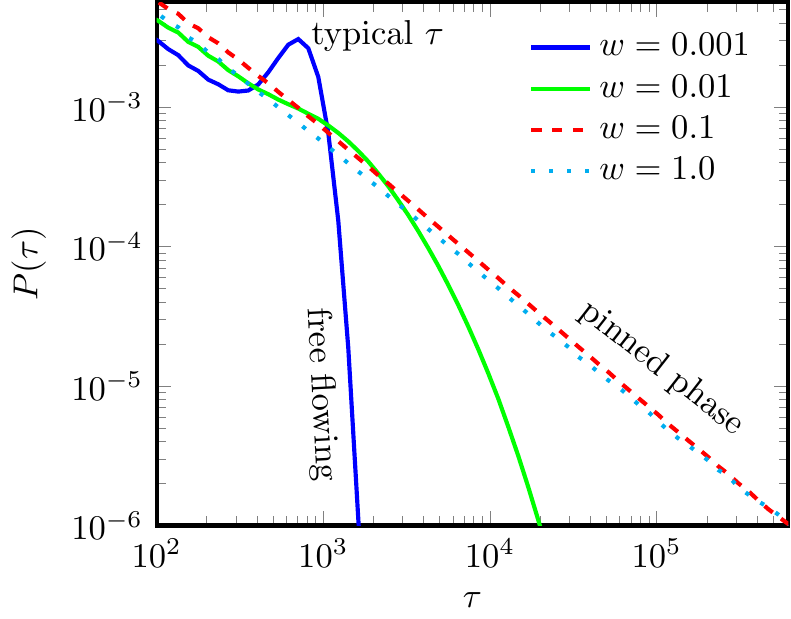} 
  \caption{Pinning time distribution $P(\tau)$ for discrete pinning sites for
    varying standard deviation $w$ of the pin strength, both in the pinned and
    moving phases, at fixed parameter values $L = 200$, 
    $T = 0.002 \epsilon_0 b_0$, $F = 0.002 \epsilon_0 b_0^{-1}$, $N_k = 100$, 
    and $\mu = 0$. The local maximum of $P(\tau)$ deep in the freely flowing 
    phase indicates a characteristic pinning time. For each parameter set, the 
    data were averaged over $1000$ independent realizations.}
  \label{fig:transition}
\end{figure}
We first explore the dynamic transition between the pinned and free-flowing 
phases of driven vortex lines at the critical depinning force $F_c$, which is 
a function of the Larkin--Ovchinnikov length $L_L$ and hence depends on defect 
potential properties, specifically the disorder strength variance $w^2$. For 
the numerical data displayed in Fig.~\ref{fig:transition}, we held the driving 
force (as well as all other parameters) fixed but varied $w$ for a system with
randomly distributed discrete point-like pins (\ref{eq:discrete_pinning}). For 
large values of the disorder strength variance, the pinning time distribution 
exhibits power law behavior. For small values of $w$, one instead observes 
$P(\tau)$ to become a quickly decaying function once $\tau > 1000 t_0$, 
indicating that the elastic lines are freely flowing. The pronounced local 
maximum visible near $\tau \approx 700 t_0$ for $w = 0.001 \epsilon_0$ marks a
characteristic average time for a line element to traverse a pinning site of 
radius $b_0$. We estimate the critical disorder variance at applied force 
$F = 0.002 \epsilon_0 b_0^{-1}$ as $w_L \approx 0.01 \epsilon_0$, wherefrom we 
may infer $\xi = w_L^{4/3} / (\epsilon^{1/3} F) \approx 1.88 b_0$. This 
disorder correlation length value matches the numerically determined mean 
nearest-neighbor distance between pinning sites $d \approx 2.7 b_0$ and the 
pinning center radius $b_0$. We have also studied samples with purely 
attractive pinning sites, i.e., truncated Gaussian disorder distributions with
width $w$ limited to $p^{(i)}_k < 0$. In the moving phase, the resulting graph 
for $P(\tau)$ coincides with the corresponding blue (dark) curve for 
$w = 0.001 \epsilon_0$ in Fig.~\ref{fig:transition}; in the pinned state, 
$P(\tau)$ is merely parallel-shifted (upwards by a factor $\approx 1.5$) from 
the dashed red line for $w = 0.1 \epsilon_0$ in Fig.~\ref{fig:transition} (data
not shown).

\subsection{Pinning time statistics for discrete disorder}
\label{sec:depinning_discrete}

\begin{figure}
  \centering
  \includegraphics{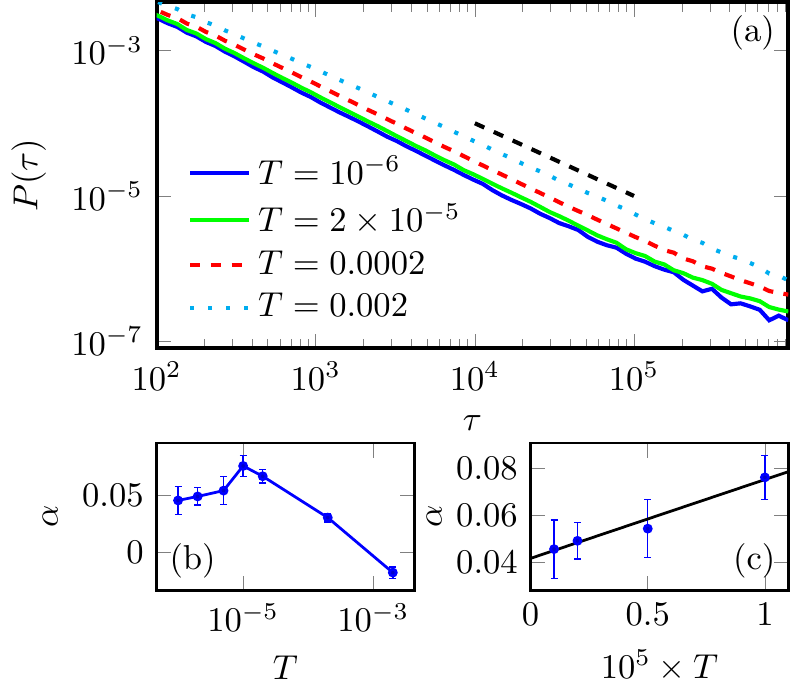}
  \caption{(a) Pinning time distribution $P(\tau)$ for discrete pinning sites 
    at different temperatures $T$ in the pinned state, with parameters 
    $L = 200$, $F = 0.002 \epsilon_0 b_0^{-1}$, $N_k = 100$, $\mu = 0$, and 
    $w = \epsilon_0$. (b) Decay exponent $\alpha(T)$ as defined in 
    Eq.~(\ref{eq:powerlaw}) as a function of temperature, obtained by fitting 
    in the region indicated by the black dashed line in (a). (c) Linear plot of 
    $\alpha(T)$ for low temperatures $T \leq 10^{-5} \epsilon_0 b_0$. The 
    solid line represents the linear fit function 
    $\alpha(T) = 0.0417 + 3340\,T$. The data were generated from $10^4$ 
    realizations for each parameter set, except for 
    $T = 0.002 \epsilon_0 b_0$, for which $1000$ realizations were used.}
  \label{fig:discrete_exponent}
\end{figure}
We next address the scaling properties of the dwelling time distribution 
$P(\tau)$ for vortex lines that are pinned to discrete localized potential 
wells (\ref{eq:discrete_pinning}) with mean strength $\mu = 0$ and variance 
$w^2$ under the influence of a subcritical driving force $F < F_c$. 
Figure~\ref{fig:discrete_exponent}(a) shows the pinning time distribution for 
$w = \epsilon_0$ and $F = 0.002 \epsilon_0 b_0^{-1}$, i.e., in the pinned 
phase, at various temperatures. Beyond $\tau \approx 10^3 t_0$, $P(\tau)$ 
settles to a power law decay over two decades. This is followed by a crossover
to a different long-time regime for $\tau > 10^5 t_0$. We have extracted the
scaling exponent $- (1 + \alpha)$ from Eq.~(\ref{eq:powerlaw}) in the 
intermediate time regime $10^4 t_0 < \tau <10^5 t_0$ by a linear fit to the
double-logarithmic data. The resulting decay exponent $\alpha(T)$ as function
of temperature is plotted in Fig.~\ref{fig:discrete_exponent}(b). This graph 
exhibits two distinct regimes: For $T \le10^{-5} \epsilon_0 b_0$ we observe
approximately linear growth, whereas $\alpha(T)$ decreases above this 
threshold and even appears to become negative close to the depinning
transition. Our data do not confirm the prediction of Ref.~\cite{Vinokur1996} 
that $\alpha \to1$ as $T$ approaches the critical depinning temperature. The 
simple power law (\ref{eq:powerlaw}) should be valid in the limit $T \to 0$ 
where the energy scales are not significantly renormalized by thermal 
fluctuations; hence one expects $\alpha(T) \propto T$ for low 
temperatures~\cite{Vinokur1996}. Figure~\ref{fig:discrete_exponent}(c) 
provides a linear plot for $\alpha(T)$ in the low-temperature region 
$T \leq 10^{-5} \epsilon_0 b_0$. We indeed find an approximately linear 
relationship with a proportionality factor $3340 (\epsilon_0 b_0)^{-1}$. 
However, the apparent nonzero intercept $\alpha(T \to 0) \approx 0.042$ is 
incompatible with Eq.~(\ref{eq:powerlaw}).

\subsection{Attractive pins versus mixed repulsive and attractive defects}
\label{sec:purely-attr-pinn}

\begin{figure}
  \centering
  \includegraphics{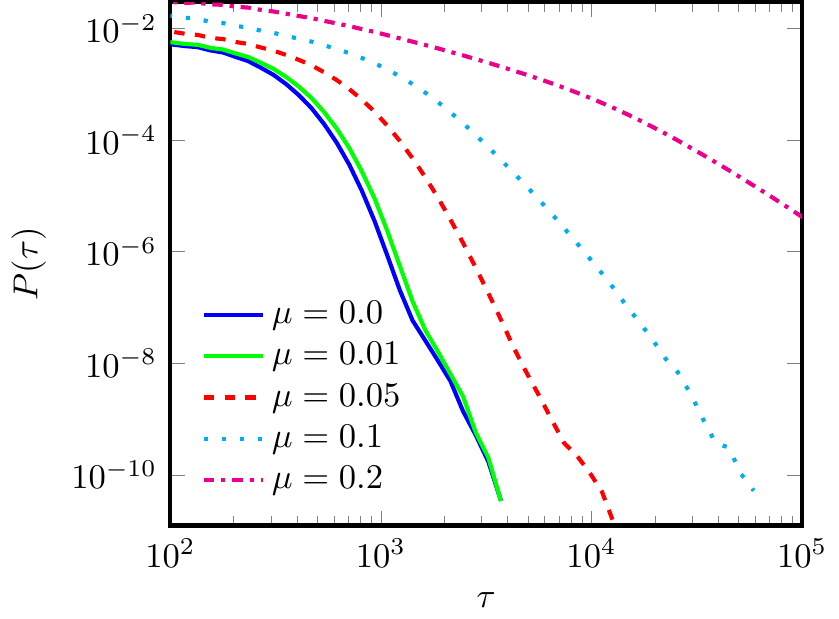} 
  \caption{Pinning time distributions $P(\tau)$ for different mean disorder
    strengths $\mu$ of discrete defect sites in a one-dimensional transverse 
    domain ($l = 3400 b_0$ and $W = 0$) for parameter values $L = 200$,
    $T = 0.02 \epsilon_0 b_0$, $F = 0.002 \epsilon_0 b_0^{-1}$, $N_k = 50$, and
    $w = 0.001 \epsilon_0$. Data in this figure were generated from a minimum
    of $6500$ realizations for each set of parameter values.}
  \label{fig:attrvsmixed}
\end{figure}
Here we wish to explore differences between largely attractive and a mixture 
of repulsive and attractive disorder. Yet in a sample with two available 
dimensions perpendicular to the directed lines, vortex segments can simply 
slide past any localized circular repulsive defects. Hence we restrict line 
movement to one perpendicular dimension in this section. 
Figure~\ref{fig:attrvsmixed} displays the pinning time distribution for 
discrete defects for various mean disorder potential strengths $\mu$ at fixed 
low standard deviation $w = 0.001 \epsilon_0$. In the symmetric case $\mu = 0$,
defect wells have an equal chance of being attractive and repulsive, while for 
finite $\mu > 0$, fewer repulsive sites are present. It is obvious from the 
graphs that upon increasing $\mu$, characteristic pinning times become longer 
due to the prevalence of deeper potential wells, and one observes a gradual 
crossover from the freely moving vortex state into the pinned phase. This 
behavior cannot be realized with a Gaussian disorder landscape, since a shift 
in the Gaussian distribution merely adds an additive constant to the 
Hamiltonian which is irrelevant for the dynamics.

\subsection{Pinning Potential Landscape}
\label{sec:landscape}

\begin{figure}
  \centering
  \includegraphics{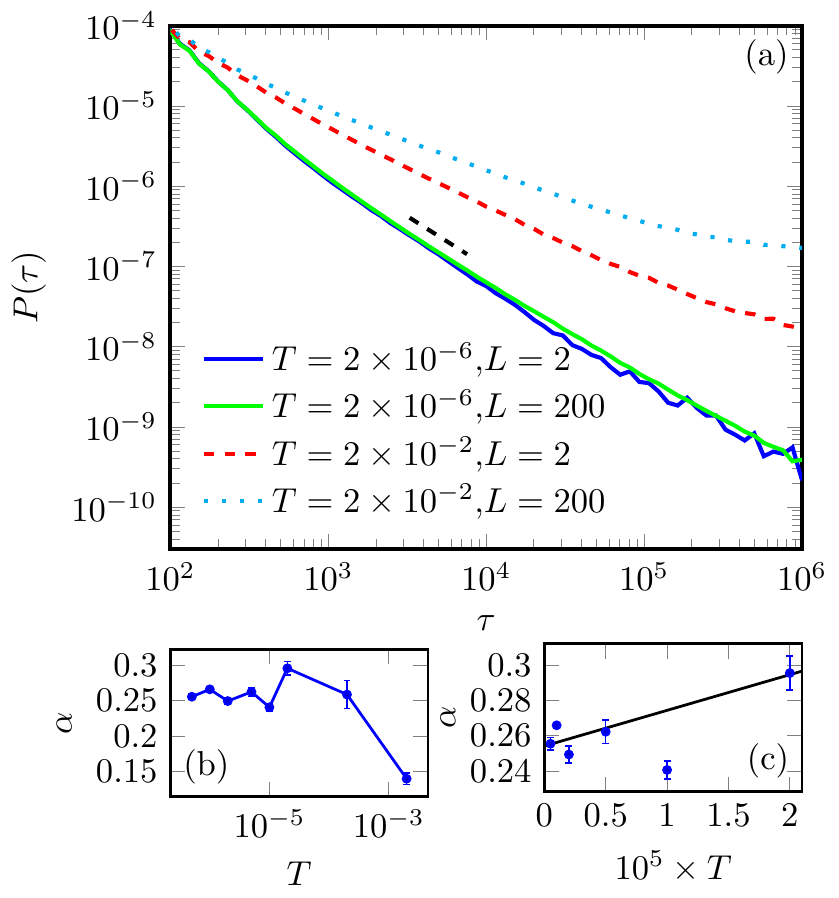} 
  \caption{(a) Pinning time distributions $P(\tau)$ for a continuous Gaussian
    disorder landscape with $\xi = b_0$ and different line lengths $L$ and
    temperatures $T$, and parameter values $F = 0.002 \epsilon_0 b_0^{-1}$ and
    $w = \epsilon_0$. (b) Decay exponent $\alpha(T)$ as function of 
    temperature for $L = 200$, obtained by fitting in the region indicated by 
    the black dashed line in (a). (c) Linear plot of $\alpha(T)$ for low 
    temperatures $T \leq 2 \times 10^{-5} \epsilon_0 b_0$. The solid line 
    represents the linear fit function $\alpha(T) = 0.2542 + 2019\,T$ 
    (excluding the outlying data point at $T = 10^{-5} \epsilon_0 b_0$). The 
    data were generated from $2 \times 10^4$ independent realizations for each 
    parameter set for $T < 2 \times 10^{-5} \epsilon_0 b_0$, else from $1000$ 
    realizations.}
\label{fig:landscape_exponent}
\end{figure}
We finally investigate the scaling properties of the dwelling time 
distribution in the pinned phase when disorder is implemented through a 
continuous Gaussian potential landscape. 
Figure~\ref{fig:landscape_exponent}(a) shows the resulting distributions 
$P(\tau)$ for different temperatures and line lengths $L = 2$ and $L = 200$, 
allowing a direct comparison with the data in Fig.~\ref{fig:discrete_exponent} 
obtained for discrete pinning sites with identical driving force 
$F = 0.002 \epsilon_0 b_0^{-1}$ and disorder strength standard deviation
$w = \epsilon_0$. Once again we observe a crossover from an early-time to the 
algebraic decay regime, but now around $\tau \approx 10^2 t_0$, indicating a
different overall energy scale. We note that at low temperatures, no 
difference in the distribution of pinning times between long and short vortex 
lines is discernible. For large temperatures, however, cooperative effects 
between the line elements come into play that effectively enhance the pinning 
times for long elastic lines. The decay exponent $\alpha(T)$ as function of 
temperature is plotted in Fig.~\ref{fig:landscape_exponent}(b), here acquired 
by fitting a power law to the data in the region indicated by the black dashed line
in (a). The decay exponent values for a continuous pinning landscale are 
significantly enhanced by a factor $\approx 4 \ldots 6$ in comparison with 
discrete pinning sites, and strictly positive for the temperature range 
investigated here. For small temperatures 
$T \leq 2 \times 10^{-5} \epsilon_0 b_0$, $\alpha(T)$ is again well described 
by a linear function of $T$, with a definitely positive zero-temperature limit 
$\alpha(T \to 0) \approx 0.254$.

\section{Conclusions}
\label{sec:conclusions}

We have employed a directed elastic line model and utilized Langevin molecular
dynamics to investigate the pinning kinetics of driven non-interacting 
magnetic vortices in type-II superconductors subject to point defects. 
We have focused specifically on the pinning time distribution $P(\tau)$ and 
its scaling properties in the pinned state. We have elucidated similarities 
and differences between distinct disorder implementations: (i) discrete 
pinning sites with localized spherical potential wells 
(\ref{eq:discrete_pinning}), whose strengths were drawn from a normal
distribution with mean $\mu$ and standard deviation $w$, and which were 
randomly distributed in the simulation domain, aside from carefully avoiding 
any spatial overlaps of the wells; and (ii) continuous Gaussian disorder 
landscapes, correlated on a length scale $\xi$ and again with variance $w^2$.

For discrete pinning sites, we found very similar dwelling time distribution
shapes for either purely attractive or symmetric defect potentials with 
$\mu = 0$ in our samples with two transverse dimensions. Restricting line
motion to merely one transverse direction, increasing the mean $\mu$ induces
a very gradual crossover from the moving to the pinned state. In contrast,
upon increasing the disorder distribution width $w$ at fixed $\mu = 0$ (as
well as driving force $F$ and temperature $T$) in the fully three-dimensional
samples, one observes a sharp transition between the freely flowing and 
pinned vortex phases. Thus we could extract a reasonable value for the 
disorder correlation length $\xi$ from the Larkin--Ovchinnikov collective 
pinning picture.
 
In the pinned phase, the dwelling time statistics for the elastic line
elements decays algebraically, cf. Eq.~(\ref{eq:powerlaw}) as predicted in 
Ref.~\cite{Vinokur1996}, with a decay exponent $\alpha(T)$ that is well 
approximated by a growing linear function at very low temperatures, but which
decreases with $T$ beyond a remarkably sharp threshold temperature. Both this 
threshold and the numerical values of $\alpha$ strongly depend on the disorder
implementation: For the continuous random potential landscape, the former is 
enhanced by about a factor $2$ for our chosen parameter values, and the latter
by roughly $5$ as compared to corresponding systems with discrete pins. We 
also observe that the zero-temperature extrapolation $\alpha(T \to 0)$ yields
a positive value, indicating perhaps subtle renormalization effects not 
entirely captured in the analysis of Ref.~\cite{Vinokur1996}.

In-depth studies of the dwelling time statistics in disordered systems thus
reveal remarkably rich features that offer novel insights in the associated
subtle physical interplay between competing energy scales.  A natural extension
of the present study would be to incorporate mutual repulsive forces between
vortex lines. Unfortunately, our currently available computing power does not
yet allow us to tackle this intriguing issue with satisfactory
statistics. Investigating the pinning time statistics for samples with spatially
correlated disorder such as columnar or planar defects is also a promising
avenue for future research. 
Because of much more efficient pinning of vortex lines in the presence of 
extended defects, we expect the associated time scales to grow and temporal
correlations to play a significant role. This renders such extended studies 
quite challenging, at least with our currently available computational 
resources. Yet we hope to be able to revisit this intriguing problem with its
many competing energy scales and associated rich dynamical properties in the 
future.

\acknowledgments{The authors wish to thank Valerii Vinokur for suggesting this
  project to us.  This research is supported by the U.S. Department of Energy,
  Office of Basic Energy Sciences, Division of Materials Sciences and
  Engineering under Award DE-FG02-09ER46613.}

\end{document}